# Comprehensive indicators and fine granularity refine density scaling laws in rural-urban systems

**Jack Sutton[1*], Quentin S. Hanley[1,2], Gerri Mortimore[3], Ovidiu Bagdasar[1,4], Haroldo V. Ribeiro[5], Thomas Peron[6], Golnaz Shahtahmassebi[7], Peter Scriven[3]**

[1]College of Science and Engineering, University of Derby, Markeaton Street, Derby, DE22 3AW, United Kingdom

[2]GH and Q Services Limited, West Studios, Sheffield Road, Chesterfield S41 7LL, United Kingdom

[3]College of Health, Psychology and Social Care, University of Derby, Kedleston Road, Derby, DE22 1GB, United Kingdom

[4]Department of Mathematics, Faculty of Exact Sciences, "1 Decembrie 1918" University of Alba Iulia, Alba Iulia, 510009, Romania

[5]Departamento de Fisica, Universidade Estadual de Maringa, Maringa, PR 87020-900, Brazil

[6]Institute of Mathematics and Computer Science, Universidade de Sao Paulo, Avenida Trabalhador Sao Carlense, 400-Centro, 13566-590 Sao Carlos, Brazil

[7]School of Science and Technology, Nottingham Trent University, Nottingham, NG11 8NS, United Kingdom

***Corresponding Author**: J.Sutton (j.sutton@derby.ac.uk)

**Abstract**

Density scaling laws complement traditional population scaling laws by enabling the analysis of the full range of human settlements and revealing rural-to-urban transitions with breakpoints at consistent population densities. However, previous studies have been constrained by the granularity of rural and urban units, as well as limitations in the quantity and diversity of indicators. This study addresses these gaps by examining Middle Layer Super Output Areas (MSOAs) in England and Wales, incorporating an extensive set of 117 indicators for the year 2021, spanning age, ethnicity, educational attainment, religion, disability, economic activity, mortality, crime, property transactions, and road accidents. Results indicate that the relationship between indicator density and population density is best described by a segmented power-law model with a consistent breakpoint (33 ± 5 persons per hectare) for 92 of the 117 indicators. Additionally, increasing granularity reveals further rural-to-urban transitions not observed at coarser spatial resolutions. Our findings also highlight the influence of population characteristics on scaling exponents, where stratifying dementia and ischaemic heart disease by older age groups (aged 70 and above) significantly affects these exponents, illustrating a protective urban effect.

# 1 Introduction

Scaling has been widely applied to better understand cities, with power laws relations with population used to describe a range of urban indicators. These studies have shown acceleration of many economic and creative measures as cities grow. The larger the city, the greater the per capita GDP, patent production, and research and development (R&D) employment [1–7]. While global urbanisation continues to rise, a significant portion of the population still lives in rural areas. Recognising the neglect of rural regions, researchers have adapted these methodologies to include these areas and incorporating a density normalisation approach [8–12]. These studies have shown that many rural-urban indicators exhibit segmentation at a similar breakpoint, highlighting fundamental differences between rural and urban regions. Mortality indicators tend to decline, while crime indicators tend to accelerate in urban regions. These differences were largely explained by age demographics, with urban areas generally characterised by a younger population, while rural regions tend to have an older demographic. These findings suggest a need to better understand demographic trends in urban regions and further demonstrate that per-capita models are only appropriate when segmentation is absent and the scaling exponent indicates linearity. Despite these limitations, per-capita models remain the dominant approach in resource allocation due to their simplicity and ease of implementation. They provide a straightforward mechanism for distributing resources based on population size. However, this method often overlooks the complex interplay between demographic characteristics and service needs, leading to potential inefficiencies and inequities. Effective resource allocation requires models that account for segmentation and non-linear scaling to better align with actual demands. For example, health system funding [13] is primarily based on per-capita assumptions. Even when adjustments have been made to account for demographic differences, these modifications have often failed to produce equitable outcomes, sometimes exacerbating disparities rather than addressing them.

This paper expands on previous research on mortality, crime, property transactions, and age demographics [9] by incorporating additional indicators and enhancing data granularity. We include road accidents alongside new demographic variables such as ethnicity, education, occupational status, religion, and disability status, bringing the total number of indicators to 117 for the year 2021. In addition, previous studies have used coarser granularity, relying on lower-tier local authorities ($n = 337$) [8], unitary authorities, non-metropolitan districts, metropolitan boroughs, and London boroughs ($n = 348$) [9], as well as parliamentary constituencies ($n = 573$) [10,11] as regional boundaries. In contrast, we examine scaling behaviour using *middle layer super output areas* (MSOA, $n = 7080$) to provide a more detailed and granular perspective for the year 2021. Finally, we stratify dementia and ischaemic heart disease by population demographic factors to evaluate their influence on scaling exponents, comparing these findings to power law models based on total population assumptions.

## 2 Methods and Materials

### 2.1 Scaling Models

Scaling models are underpinned by power laws, where an urban indicator $y$ is a function of population $x$. In the standard urban scaling model, the power law is given by:

$$y = \alpha x^\beta \varepsilon \tag{1}$$

where $\alpha$ is the pre-exponential factor, $\beta$ is the scaling exponent and $\varepsilon$ represents the residuals. Model parameters can be estimated using the least squares method by linearising Equation (1) through a logarithmic transformation, as follows:

$$\log_{10}(y) = \log_{10}(\alpha) + \beta \log_{10}(x) + \log_{10}(\varepsilon) \tag{2}$$

where $\log_{10}(y)$ is the dependent variable and $\log_{10}(x)$ is the independent variable. The scaling exponent $\beta$ indicates different types of scaling: sublinear ($\beta < 1$), linear ($\beta = 1$) and superlinear ($\beta > 1$).

To incorporate the rural-urban framework, Equations (1-2) are adjusted from an indicator to a density indicator $y_d$, defined as $y_d = y/A$, where $A$ is the land area of an MSOA region. This density indicator then becomes a function of population density, $x_d = x/A$, as follows:

$$y_d = \alpha_d x_d^{\beta_d} \varepsilon \tag{3}$$

where $\alpha_d$ is the density pre-exponential factor and $\beta_d$ is the density scaling exponent. Since both $y$ and $x$ are normalized by the same area factor $A$, this transformation only modifies the pre-exponential factor, while the scaling exponent $\beta_d$ remains unchanged.

Similarly, applying the logarithmic transformation linearises Equation (3), allowing the use of the least squares method. The corresponding logarithmic form is as follows:

$$\log_{10}(y_d) = \log_{10}(\alpha_d) + \beta_d \log_{10}(x_d) + \log_{10}(\varepsilon) \tag{4}$$

Similar to Equations (1-2), the density scaling exponent $\beta_d$ indicates the type of density scaling: sublinear when $\beta_d < 1$, linear when $\beta_d = 1$ and superlinear when $\beta_d > 1$.

Studies have shown that many metrics exhibit a breakpoint $\log_{10}(d^*)$ within a consistent range of 10-70 people per hectare [9,11]. To account for this transition, Equation (4) is adjusted to allow for segmentation, as follows:

$$\log_{10}(y_d) = \begin{cases} \log_{10}(y_0) + \beta_L \log_{10}(x_d) + \log_{10}(\varepsilon) & x_d < \log_{10}(d^*) \\ \log_{10}(y_1) + \beta_H \log_{10}(x_d) + \log_{10}(\varepsilon) & x_d \geq \log_{10}(d^*) \end{cases} \quad (5)$$

where $\log_{10}(y_0)$ and $\beta_L$ denote the density pre-exponential factor and density scaling exponent below the break points $\log_{10}(d^*)$, while $\log_{10}(y_1)$ and $\beta_H$ are the density pre-exponential factor and density scaling exponent above the break point $\log_{10}(d^*)$. We fit both single and segmented models to explore the relationship between indicator density and population density, identifying the best fit model using Davies' test with 99% confidence, as well as the Akaike (AIC) and Bayesian (BIC) Information Criterion. After fitting the model, either using a single or segmented power law, residuals are obtained as follows:

$$\log_{10}(\varepsilon_i) = \log_{10}(y_{d,i}) - \log_{10}(\hat{y}_{d,i}) \quad (6)$$

where $i = 1, \ldots, n$, with $n$ representing the number of MSOA regions and $\log_{10}(\hat{y}_{d,i})$ is the estimate of $\log_{10}(y_{d,i})$. Since standard linear regression methods are applied, it is assumed that the residuals $\varepsilon$ follow a normal distribution, with $\log_{10}(\varepsilon)$ assumed to be independent and identically distributed (i.i.d.) as $N(0, \sigma^2)$. Model parameters are estimated using the least squares method applied to the logarithmic transformations (i.e. $\log_{10}(y)$ vs. $\log_{10}(x)$), which aims to minimise the sum of squared residuals, $\sum (\log_{10}(\hat{\varepsilon}_i))^2$.

## 2.2 Data, Software and Packages

Data on mortality, age, ethnicity, education, economic activity status, religion, and disability for England and Wales were provided by NOMIS (https://www.nomisweb.co.uk/). NOMIS anonymise mortality data by adjusting low counts so that values $\leq 2$ are set to 0 and values between 3 and 5 are set to 5. Data on age, ethnicity, education, economic activity, and disability are based on the most recent census (March 2021) and estimated based on residency in England and Wales. Population, land area, crime, property value transactions and road accidents data were obtained from UKCrimeStats (https://www.ukcrimestats.com), which sources these statistics from public platforms such as the Home Office, and the Land

Registry and aligns them using geographical shape files obtained from the Ordnance Survey Boundary Line (https://www.ordnancesurvey.co.uk/).

We collected 117 indicators including data for mortality ($n_{\text{mortality}} = 41$), crime ($n_{\text{crime}} = 14$), property transactions ($n_{\text{property}} = 9$) and road accidents ($n_{\text{road}} = 1$) for 2021. The remaining variables correspond to population characteristics, including age categories ($n_{\text{age}} = 18$), ethnic groups ($n_{\text{ethnicity}} = 5$), qualification levels ($n_{\text{education}} = 7$), economic activity status ($n_{\text{economic}} = 9$), religious groups ($n_{\text{religion}} = 9$) and disability status ($n_{\text{disability}} = 4$), based on estimates from the 2021 census. A comprehensive list of indicators (Table S1) along with the complete formatted and aligned data (Dataset S1) are available in the supplementary material. Data related to dementia and ischaemic heart disease were stratified by age groups and are also available in the supplementary material (Dataset S2-3).

Statistical modelling and data analysis were completed using R (version 4.1.2) [14] and R Studio (version 2021.09.0). Data Loading, writing and formatting were handled with the xlsx (version 0.6.5) [15] and rio (version 0.5.27) [16] packages. Segmented power law models were fitted using the segmented (version 1.1-1) [17–21] package and exponents presented using the gplots (version 3.1.3) [22] package.

## 3 Results

### 3.1 Scaling with improved granularity

England and Wales consist of 7,080 MSOA regions. The area of these regions ranged from 29.42 ha (Hammersmith and Fulham 007: E02000378) to 111,715.9 ha (Northumberland 019: E02005727). Population sizes ranged from 2,226 (Isles of Scilly: E02006781) to 23,504 (Sheffield: E02006843), while population density ranged from 0.06 people per hectare (Northumberland 019: E02005727) to 294.53 people per hectare (Westminster 022: E02000981). Density scaling models were fitted to all indicators (e.g. Figures 1-3; Figures S1-9). Both single (equation 4) and segmented (equation 5) density models were evaluated using the Davies test, AIC and BIC scores to identify the best fit. Complete results are available in the supplementary material (S4 Dataset).

Most indicators gave reasonable fits to single or segmented power laws (Figures 1-3; S1-S9) providing a localised perspective of scale. Overall, regions did not stand out relative to the power laws with the notable exception of violent crimes and to a lesser extent: anti-social behaviour (ASB), criminal damage and arson (CD&A) and order. For example, violent crime showed significant negative deviations from the power law relation for regions of Greater Manchester (Figure 1). This behaviour has not been observed previously and did not appear in all other crime indicators. Issues with the Greater Manchester Police Constabulary responsible for these regions have been noted previously indicating under-reporting of crime [23,24]. The data here suggest under-reporting is confined to specific types of crime and does not arise from a process affecting all types of crime. The behaviour seen highlights the opportunity provided by population density based scaling methods at MSOA granularity to identify local issues. It is also notable that the under-reporting MSOAs result in a parallel relationship and as such will have little or no effect on the scaling exponent.

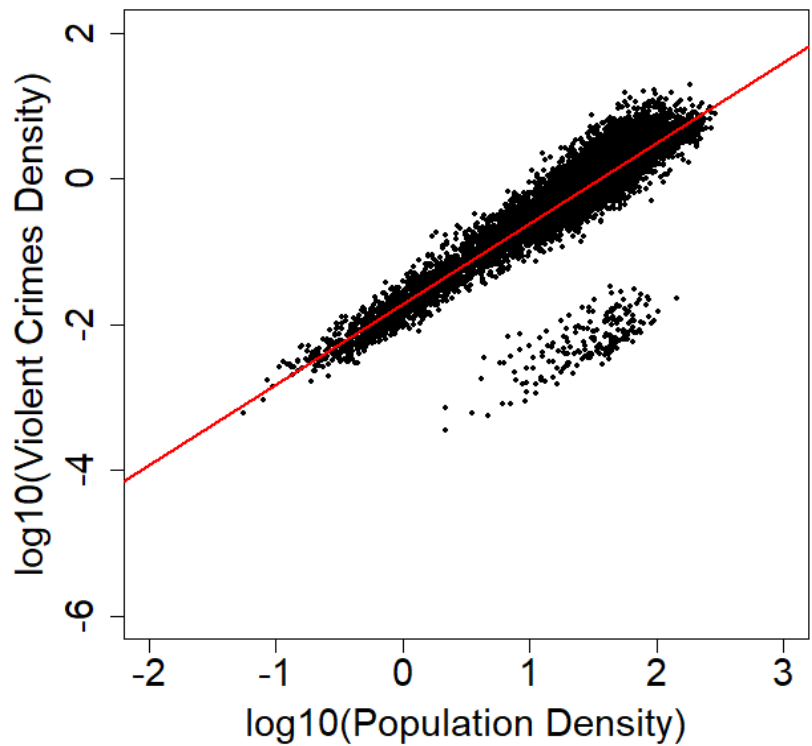

Figure 1. Density scaling plot for violent crimes in 2021. Black dots represent MSOA regions. When single power law applies, the solid red line illustrates the fitted model across the entire range of population densities.

An examination of crime types, road accidents (Figure 2, panel A) and property transactions (Figure 2, panel B) reveals all, with the exception of violence (includes sexual offenses), were better fitted with a segmented power law. In previous studies with less granularity, anti-social behaviour (ASB), other crime, burglary, weapons, order, vehicle crime, bike theft, and freehold properties were better fitted with a single power law. The more granular MSOA data suggests these scaling behaviours are better captured by a segmented power law indicating that aggregation of larger areas masks some features of scaling. The average breakpoint $d^*$ across all segmented indicators was 33 people per hectare (p/h). This was broadly in agreement with 30 p/h for 573 parliamentary constituencies [11] and 27 p/h for 348 regions [9]. The process creating this critical boundary between rural and urban areas is remarkably robust across multiple data sets and a wide range of indicators. With increased granularity additional breakpoints were revealed with possibly a slight tendency toward increasing $d^*$.
.

A total of 41 mortality related metrics were studied. This was augmented with a safety related non-mortality metric, road accidents (Figure 2, panel D) which exhibited segmented scaling changing from sublinear to superlinear across the breakpoint. This disproportionate increase in road accidents with increasing population density may be due to increased traffic, complex road networks, and more interactions between pedestrians, cyclists, and vehicles. This is supported by studies showing that accidents increase superlinearly with population size, particularly with lower severity in urban areas compared to more fatal accidents in rural areas[25]. Approximately half of mortalities (17 out of 41) exhibited segmented scaling with exponents declining in the highly urbanised regions above $d^*$ (Figure 2 panel C). This has been noted previously and has been explained by age demographics. The remaining mortalities (24 out of 41) were better fitted by a single scaling model, often involving rare diseases or regions with limited mortality reporting. We ascribe this to these diseases not being associated with environmental factors or lifestyle in any way. It is worth noting that at this granular level, zero values become problematic, as extensively discussed in the scaling literature and remains an active area of research[26].

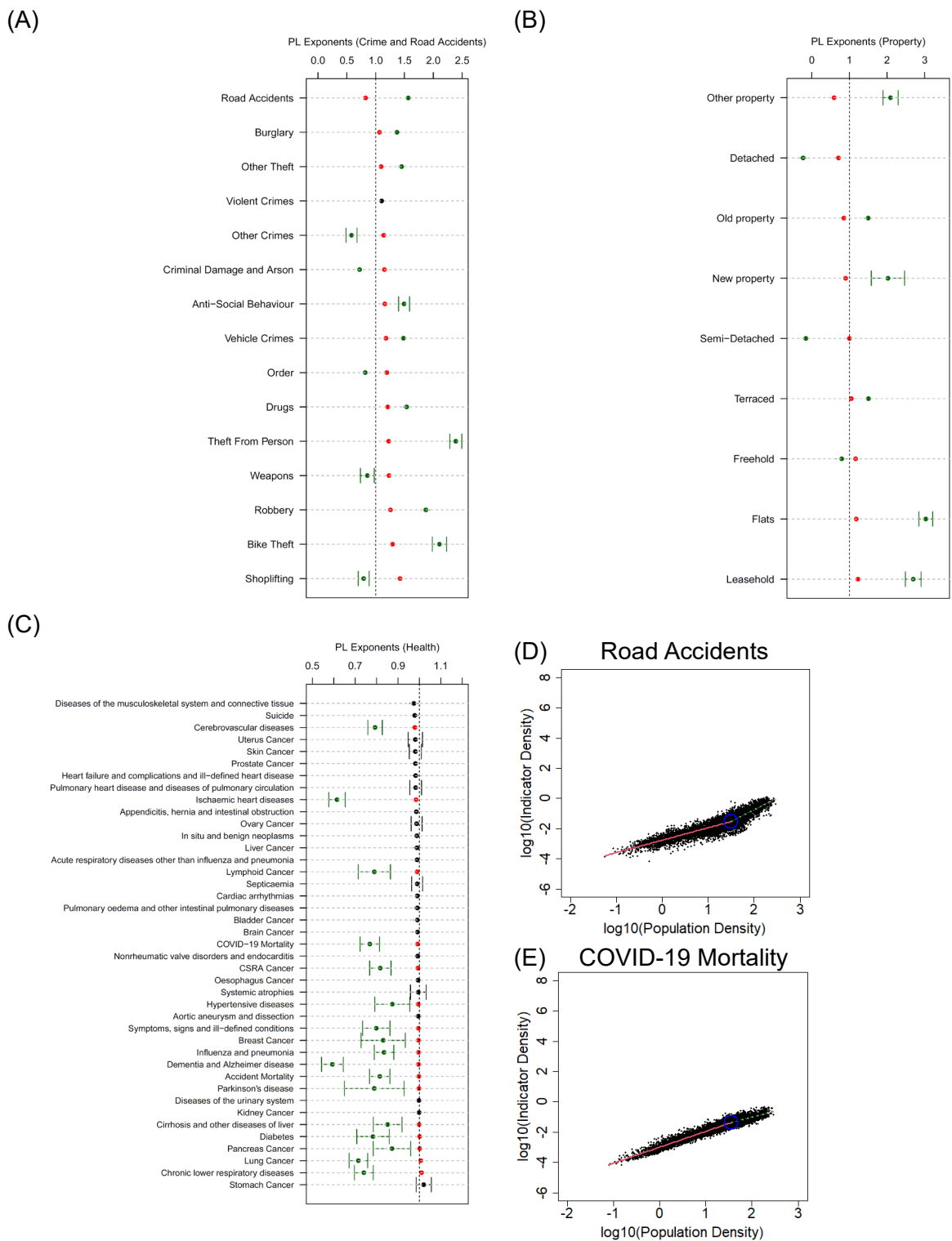

Figure 2. Density scaling exponents for (A) road accidents and crime, (B) property transactions and (C) mortality. In panels (A-C) red dots are reported density scaling exponents ($\beta_L$) below the breakpoint $\log_{10}(d^*)$, while green dots are reported density scaling exponents ($\beta_H$) above the breakpoint $\log_{10}(d^*)$. The vertical dashed black line represents linear scaling; point to the left of the line indicate sublinear scaling, and points to the right indicate superlinear scaling. Error bars represent the 95% confidence intervals for $\beta_L$ and $\beta_H$ based on the standard errors of regression. If error bars are not present the errors are very small. Panels (D-E) are examples of density scaling plots. These are (D) road accidents and (E) COVID-19 mortality. In panels (D-E) black dots represent MSOA regions, with the blue circle indicating the position of the breakpoint. The red line ($\beta_L$), represents the expectation relative to population density below the breakpoint, while the green dashed line ($\beta_H$), represents the expectation relative to population density above the breakpoint.

## 3.2 Additional Population Characteristics

Previously, the scaling of specific age groups was shown to scale variably with population density in the UK. Younger people appeared to be attracted to urban environments while older people seemed to leave. This helped explained acceleration in crime and reduced mortality in urban areas, both of which are expected in a younger population. An examination of the scaling of additional population characteristics revealed considerable additional information unobservable if total population is used. This analysis included: age groups (Figure 3, panel A), ethnicity (Figure 3, panel B), education (Figure 3, panel C), occupation status (Figure 3, panel D), religion (Figure 3, panel E), and disability status (Figure 3, panel F).

Consistent with previous work [9], scaling exponents varied by age group such that young adults (20 to 44) accumulate and people aged 45 and older decline in more urban environments. The urban sublinear trend observed in older age groups tends to be more pronounced with age. With one exception, ethnic groups (Asian, Black, other group, and mixed or multiple ethnic groups) showed superlinear scaling with acceleration observed beyond the breakpoint in highly dense regions. People identifying as white ethnic group displayed linear to sublinear scaling, suggesting urban decline in highly dense regions (Figure 3 panel C).

Cities serve as central hubs for education leading to Level 4 and above qualifications exhibiting urban acceleration (Figure 3, panel D). Levels 4 to 6 roughly correspond to years 1 to 3 of a 3-year undergraduate degree (e.g., BSc). Beyond this are level 7 (e.g. Master's degrees) and 8 (e.g., PhD). A range of other professional and other qualifications (e.g., teaching) fit into this scheme. Most of these qualifications are primarily provided by universities which are predominantly located in cities; therefore, urban university infrastructure and the younger age demographic they attract help explain this urban superlinearity in educational attainment [27]. By contrast, Level 1–3 qualifications, apprenticeships and other qualifications covering a wide range of academic achievement as well as no qualifications exhibit linear to sublinear scaling, indicative of economies of scale.

Almost all economic status (Figure 3 panel D) indicators exhibit segmented with a transition from linear to superlinear scaling. The notable exceptions are retired people and those with long term sickness or disability which indicate reduction in more urban areas. This can largely be explained by age demographics. The rural side of the breakpoint tends to be older and older people are more likely to fall into these categories. This also relates to the scaling observed in disability indicators (Figure 3 panel E), where low disability, high disability, and long-term physical or mental health conditions (LTPMC) exhibit linear to sublinear scaling.

Differences in scaling behaviours was also seen across religious groups. Two-thirds of religious groups (Other religion, Not answered, Jewish, Buddhist, Hindu, and Muslim) exhibited urban acceleration above the power law breakpoint, while the remaining groups (Christian, No religion, and Sikh) showed urban decline above the breakpoint. Overall, the differential scaling behaviour observed for ethnic and religious groups, economic activity status, long-term illness and disability challenges the notion that the vibrance of cities can be understood simply by the total population. Total population analysis ignores the accumulation of vibrant, young, well-educated, healthy people in urban areas.

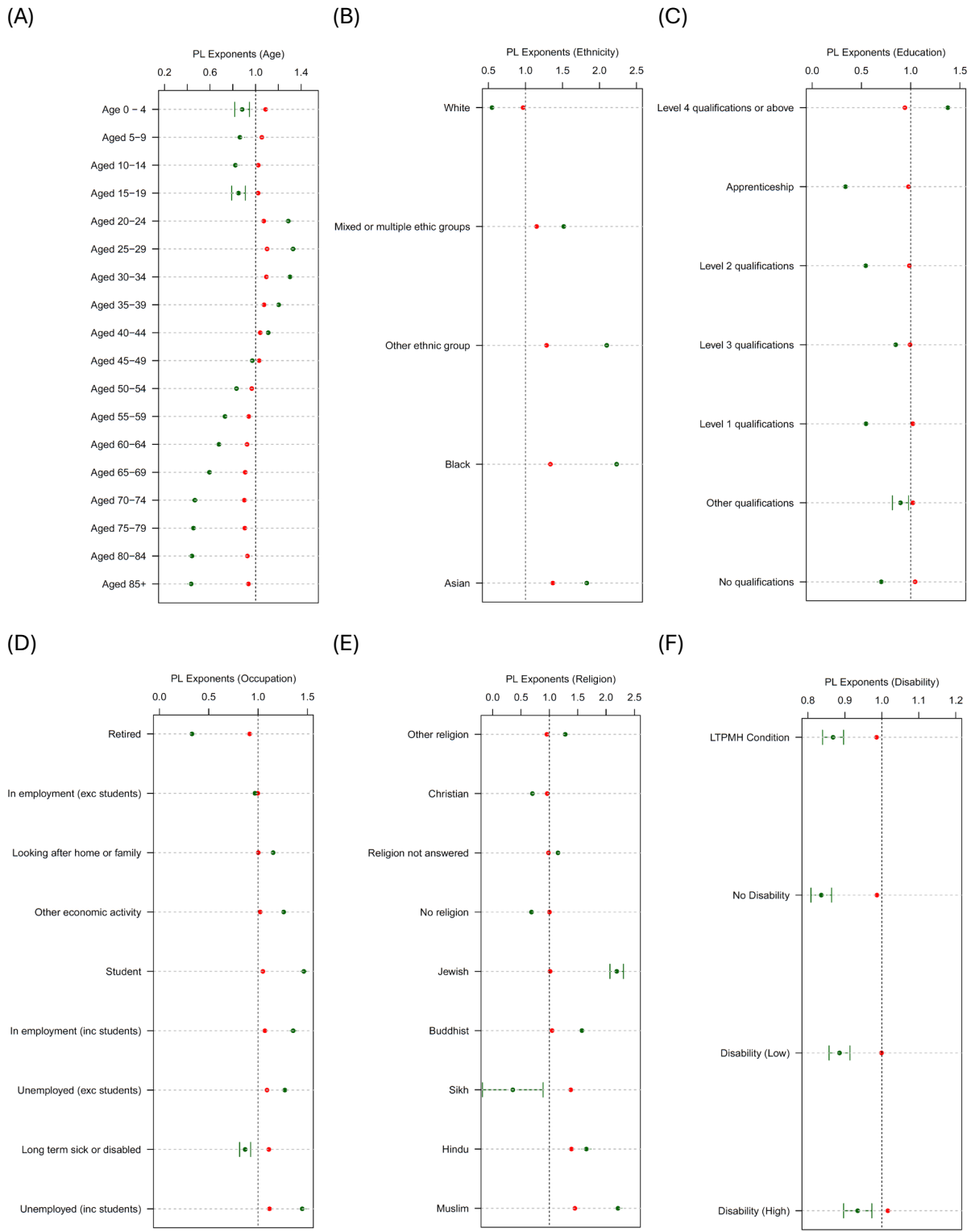

Figure 3. Density scaling exponents for population characteristics. In panels (A-F) red dots are reported density scaling exponents ($\beta_L$) below the breakpoint $\log_{10}(d^*)$, while green dots are reported density scaling exponents ($\beta_H$) above the breakpoint $\log_{10}(d^*)$. The vertical dashed black line represents linear scaling; point to the left of the line indicate sublinear scaling, and points to the right indicate superlinear scaling. Error bars represent the 95% confidence intervals for $\beta_L$ and $\beta_H$ based on the standard errors of regression. If error bars are not present the errors are very small.

## 3.3 Demographics and Mortality Scaling Exponents

To better understand the impact of population characteristics on scaling exponents, we restricted the total population by stratifying mortality by older age groups (Tables 1–2, Figures 4–5). We then compared scaling exponents across older demographic groups and examined how they differ when considering the general population.

As shown in Table 1 and Figure 4, rural scaling exponents below the consistent breakpoint for dementia remain relatively consistent across older age groups, indicating linear scaling behaviour. In contrast, urban scaling exponents above the breakpoint for individuals aged 75 and over exhibit sublinear scaling, exhibiting an urban protective effect against dementia. This protective effect appears to extend to ischaemic heart disease (Table 2, Figure 5), where a similar pattern emerges. Above a consistent breakpoint, individuals aged 75 and over also display sublinear scaling.

The underlying demographics of a region determines its behaviour. From a policy and decision-making perspective, these population characteristics are crucial considerations. Resource allocation or policies that rely solely on general population and per-capita assumptions provide an oversimplified view of the complexities involved in understanding population at scale. This can lead to underestimation, resulting in higher costs, increased burden and negative outcomes. It is possible that these urban protective effects extend to other population characteristics, such as attainment and occupation status, but understanding this depends on the availability of stratified data.

| Dementia stratified by | $n$ | Single exponent, $\beta_d$ | Lower segmented exponent, $\beta_L$ | Upper segmented exponent, $\beta_H$ | Breakpoint, $\log_{10}(d^*)$ |
|---|---|---|---|---|---|
| Aged 70-74 | 69 | 1.09 [1.00:1.18] | - | - | - |
| Aged 75-79 | 460 | - | 1.12 [0.87:1.38] | 0.72 [0.47:0.98] | 0.21 [0.07:0.36] |
| Aged 80-84 | 1505 | - | 1.08 [0.95:1.21] | 0.83 [0.70:0.95] | -0.06 [-0.17:0.06] |
| Aged 85 + | 4213 | - | 1.05 [0.98:1.13] | 0.87 [0.80:0.94] | -0.12 [-0.24:0.00] |

Table 1. Dementia stratified by older age groups, showing single exponents and, where applicable, lower and upper exponents along with breakpoints.

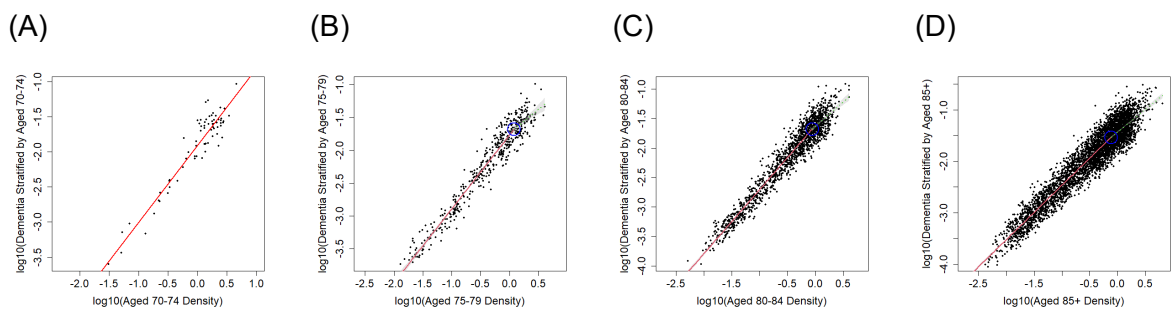

(A) (B) (C) (D)

Figure 4. Density scaling plots for dementia in 2021 stratified by older age groups. Black dots represent MSOA regions. In the case of segmented behaviour, blue circles indicate the position of the breakpoint ($\log_{10}(d^*)$) where the red line ($\beta_L$), represents the expectation relative to population density below the breakpoint ($\log_{10}(d^*)$), while the green dashed line ($\beta_H$), represents the expectation relative to population density above the breakpoint ($\log_{10}(d^*)$). When single power law applies, the solid red line illustrates the fitted model across the entire range of population densities.

| Heart disease stratified by | $n$ | Single exponent, $\beta_d$ | Lower segmented exponent, $\beta_L$ | Upper segmented exponent, $\beta_H$ | Breakpoint, $\log_{10}(d^*)$ |
|---|---|---|---|---|---|
| Aged 55-59 | 70 | 1.06 [1.00:1.11] | - | - | - |
| Aged 60-64 | 164 | 1.09 [1.06:1.12] | - | - | - |
| Aged 65-69 | 265 | 1.10 [1.08:1.13] | - | - | - |
| Aged 70-74 | 615 | 1.11 [1.09:1.13] | - | - | - |
| Aged 75-79 | 789 | - | 1.11 [0.92:1.30] | 0.77 [0.57:0.96] | 0.10 [-0.01:0.22] |
| Aged 80-84 | 1020 | - | 1.05 [0.88:1.23] | 0.79 [0.62:0.96] | 0.01 [-0.12:0.14] |
| Aged 85 + | 1836 | - | 1.03 [0.90:1.16] | 0.67 [0.54:0.80] | -0.02 [-0.10:0.06] |

Table 2. Ischaemic heart diseases stratified by older age groups, showing single exponents and, where applicable, lower and upper exponents along with breakpoints.

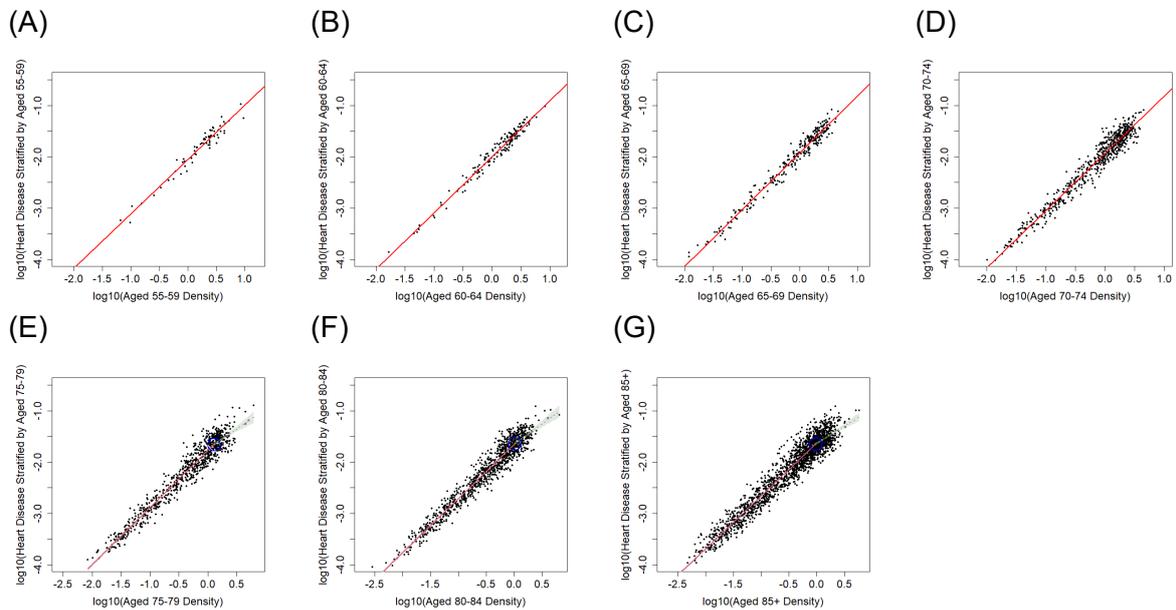

Figure 5. Density scaling plots for Ischaemic heart diseases in 2021 stratified by older age groups. Black dots represent MSOA regions. In the case of segmented behaviour, blue circles indicate the position of the breakpoint ($\log_{10}(d^*)$) where the red line ($\beta_L$), represents the expectation relative to population density below the breakpoint ($\log_{10}(d^*)$), while the green dashed line ($\beta_H$), represents the expectation relative to population density above the breakpoint ($\log_{10}(d^*)$). When single power law applies, the solid red line illustrates the fitted model across the entire range of population densities.

## 4 Discussion

This study investigated the influence of data granularity and population demographics on scaling. Increasing the number of areas describing England and Wales to 7080 from 348 allowed segmented scaling to be observed for some indicators. There are limits to increasing granularity, particularly for rare events since the values in smaller regions may be insufficient for meaningful analysis; however, the finer spatial resolution uncovered previously masked transitions, such as the superlinear escalation of burglary, other theft, and several other crime types in high density regions. The high granularity data was able to independently diagnose under-reporting in a police constabulary affecting only a selection of crimes. Future work on monthly data could help assess progress toward rectifying this problem. In both cases, increased granularity revealed features (break points and underreporting) that otherwise would have been hidden. The inclusion of population characteristics highlights the relationship between demographics and scaling exponents.

Vastly different behaviours are seen in mortality depending on which demographic is used. This applies to age, ethnicity, education, religion, and economic activity classification. Classical urban scaling analysis assumes that total population size determines scaling behaviour and neglects rural regions entirely. Here, this view has been comprehensively overturned. We find that urban regions are moderately protective of people aged 75+ against mortality from dementia and ischaemic heart diseases. The unique character of urban regions rests on critical demographics within them, not on total population. These insights emphasise the need for a more nuanced approach to service planning, resource allocation, and policy design, ensuring that socio-economic factors and contextual differences are adequately considered in decision-making.

## Data availability statement

All data generated or analysed during this study are included in this published article (and its supplementary information files). This data was compiled from a range of publicly available sources as noted in the manuscript. These are provided as the Following files: S1Dataset.csv, S2Dataset.csv, S3Dataset.csv and S4Dataset.csv.

## Code availability statement

We have also provided a set of R-scripts as supplementary information. This has been provided as S1Code.R.

## Acknowledgements

This research was funded by NHS England through its Secure Data Environment initiative. Additional support was provided by '1 Decembrie 1918' University of Alba Iulia through scientific research funds. To ensure Open Access, the authors have applied a Creative Commons Attribution (CC BY) license to any Author Accepted Manuscript version resulting from this submission.

## Author Contributions

J.S., Q.S.H., G.M., O.B., H.V.R., T.P., G.S., and P.S. designed research, performed research, analyzed data and wrote the paper.

## Competing interests

The authors declare no competing interests

## Additional information

**Supplementary information** is available for this paper

**Correspondence** and requests for material should be addressed to J.S.